\documentclass[12pt]{article}
\normalbaselines \topmargin -0.25 truein \oddsidemargin 0.30
truein \evensidemargin 0.30 truein 

\font\tbf = cmbx12

\begin{document}

\begin{center}
{\bf KINETIC THEORY OF NON-ABELIAN PLASMA:  \\ A NON-MINIMAL
MODEL}

\end{center}


\centerline{\tbf Alexander Balakin\footnote{e-mail:
Alexander.Balakin@ksu.ru}  }

\vskip 0.3 cm

\centerline{\it Kazan State University}

\centerline{\it Kremlevskaya street 18, 420008, Kazan, Russia}

\begin{abstract}

The self-consistent system of master equations describing the
kinetics of a relativistic non-Abelian plasma, influenced by
curvature interactions, is formulated. Non-minimal (curvature
induced) coupling is shown to modify all the subsystems of the
model: the gauge field equations, the gravity field equations and
the kinetic equation for colored particles.

\end{abstract}

\section{Introduction}

Relativistic kinetic theory of non-Abelian plasma in its {\it
minimal} version is developed in detail (see, e.g., reviews
\cite{Else,Litim} for references). This theory is a direct
non-Abelian generalization of the kinetic theory of electrically
charged relativistic particles and focuses on the problems of
quark-gluon plasma dynamics \cite{Hwa}. Taking into account
numerous astrophysical and cosmological applications of the
quark-gluon plasma theory, it is reasonable to discuss the {\it
non-minimal} extension of the model describing colored
multi-particle systems in a strong gravitational field.
Non-minimal extension means that the Riemann, Ricci tensors and
Ricci scalar appear in the kinetic equations, in the gauge field
equations and in the gravity field equations as extra terms,
vanishing in the absence of tidal (curvature induced) interactions
(see, e.g., \cite{FaraR,Hehl3}). The non-minimal extension of the
gauge field equations was discussed in many papers, and a number
of approaches have been proposed. One of them is connected with a
dimensional reduction of the Gauss-Bonnet action \cite{MH}. We use
a non-Abelian generalization of the non-minimal Einstein-Maxwell
theory  along the lines proposed by Drummond and Hathrell for the
linear electrodynamics \cite{Drum}. Based on the results of the
paper \cite{BL05}, we considered in \cite{1BZ06,2BZ06,BSZ07} a
three-parameter gauge-invariant non-minimal Einstein-Yang-Mills
model, linear in curvature. The same idea is used here to form the
subsystem of non-minimally extended gauge field equations and
gravity field equations. The generalization of the kinetic
equation for the non-Abelian plasma, presented here, is based on
three key elements. The first one is the theory of extended phase
space (see, e.g.,
\cite{Laptev,Vlasov,Isr78,FT80,Else,Litim,Krasnoyarsk}). The
second element is a four-vector of the (non-minimal)
self-interaction force, changing the momentum of a colored
particle. This force enters the equation of particle dynamics and
is a subject of phenomenological modeling. The third element is
the scalar (non-minimal) force-like term, which predetermines the
evolution of color charges. This quantity appears in the equation
of charge dynamics, the generalization of the Wong equation
\cite{Wong}, and is modelled here phenomenologically.

\section{Kinetic equation on the extended phase space}

\subsection{Distribution function}

The extended phase space in the non-minimal non-Abelian
($SU(n)$-symmetric) kinetic theory is $(7+ n^2)$-dimensional,
i.e., is based on four spacetime coordinates $x^i$, four-vector of
particle momentum $p^k$ and $n^2-1$ color charges $Q^{(a)}$
treated as stochastic scalar variables, the distribution function
being of the form
\begin{equation}\label{distrib}
f\left(x^i,p^k, Q^{(a)} \right) \delta \left(p_n p^n {-}m^2c^2
\right) \delta \left(G_{(a)(b)}Q^{(a)}Q^{(b)}{-} \alpha_2 \right)
\delta \left(d_{(a)(b)(c)}Q^{(a)}Q^{(b)}Q^{(c)}{-} \alpha_3
\right)...
\end{equation}
The Latin indices in parentheses stand for the group ones taking
$n^2-1$ values, the symmetric tensor $G_{(a)(b)}$, defined as
\begin{equation}
G_{(a)(b)} \equiv 2 {\rm Tr} \ {\bf t}_{(a)} {\bf t}_{(b)} \,,
\label{scalarproduct}
\end{equation}
plays a role of a metric in the group space. ${\bf t}_{(a)}$ are
the Hermitian traceless generators of $SU(n)$ group, they satisfy
the relations
\begin{equation}
\left[ {\bf t}_{(a)} , {\bf t}_{(b)} \right] = i  f^{(c)}_{\
(a)(b)} {\bf t}_{(c)} \,, \label{fabc}
\end{equation}
where $f^{(c)}_{\ (a)(b)}$ are the structure constants of the
$SU(n)$ group. The first delta-function in (\ref{distrib})
guaranties that the particle momentum lies on the mass-shell. The
second one prescribes the square of color charges to be constant
and equal to $\alpha_2$. When we deal with $SU(n)$ group and
$n>2$, the third order Casimir invariant,
$d_{(a)(b)(c)}Q^{(a)}Q^{(b)}Q^{(c)}$, has to be constant, and the
third delta-function appears in (\ref{distrib}),.. etc. The
completely symmetric coefficients $d_{(a)(b)(c)}$ are defined by
\begin{equation}
\left\{ {\bf t}_{(a)} , {\bf t}_{(b)} \right\} \equiv {\bf
t}_{(a)}  {\bf t}_{(b)} + {\bf t}_{(b)} {\bf t}_{(a)} =
\frac{1}{n} \delta_{(a)(b)} {\bf I} + d^{(c)}_{\ (a)(b)} {\bf
t}_{(c)} \,, \label{dabc}
\end{equation}
${\bf I}$ being the matrix-unity. Using the complete and
incomplete measures
\begin{equation}\label{dQ}
d {\cal Q} \equiv d Q^{(1)} d Q^{(2)} ... d Q^{(n^2-1)} \,, \quad
d {\cal Q} / dQ^{(b)} \equiv d Q^{(1)}... d Q^{(b-1)} d Q^{(b+1)}
d Q^{(n^2-1)} \,,
\end{equation}
one can define, respectively, the averaged and specific
distribution functions:
\begin{equation}\label{Phi}
f\left(x^i,p^k \right) {=} \int d{\cal Q} f \left(x^i,p^k,
Q^{(a)}\right), \quad f_{(b)}\left(x^i,p^k, Q^{(b)} \right) {=}
\int d{\cal Q} / dQ^{(b)} f \left(x^i,p^k, Q^{(a)}\right).
\end{equation}

\subsection{Kinetic equation and its characteristics}

The distribution function is considered to be a solution of the
kinetic equation
\begin{equation}\label{GRKE}
\frac{p^i}{mc}\widehat\nabla_i f +\frac{\partial}{\partial{p^i}}
\left({\cal F}^i f \right) + \frac{\partial}{\partial{Q^{(a)}}}
\left({\cal H}^{(a)} f \right) = {\cal J }\,.
\end{equation}
Here $\widehat\nabla_i$ is the Cartan derivative \cite{Vlasov}
\begin{equation}\label{CartanPro}
\widehat\nabla_i \equiv \nabla_i -\Gamma^k_{il}p^l
\frac{\partial}{\partial{p^k}} \,, \quad \widehat\nabla_i p^k = 0
\,, \quad \widehat\nabla_i Q^{(a)} =0 \,,
\end{equation}
$\nabla_i$ is the covariant derivative and $\Gamma^k_{il}$ are the
Christoffel symbols, associated with the spacetime metric
$g_{ik}$. ${\cal J }$ symbolizes the collision integral, the
four-vector ${\cal F}^i$ is a generalized force, depending on
coordinates, particle momentum and color charges, the scalar
function ${\cal H}^{(a)}$ is a force-like term predetermining the
color charge evolution. The physical sense of the terms ${\cal
F}^i$ and ${\cal H}^{(a)}$ can be clarified using the
characteristic equations for the kinetic equation (\ref{GRKE}):
\begin{equation}\label{CharEq1}
\frac{dx^i}{ds}=\frac{p^i}{mc}\,, \quad \frac{dp^i}{ds}+
\frac{1}{mc}\Gamma^i_{kl} \ p^k p^l = {\cal F}^i\,, \quad
\frac{d}{ds} Q^{(a)}={\cal H}^{(a)} \,.
\end{equation}
As a consequence of (\ref{distrib}), the force ${\cal F}^i$ is
orthogonal to the particle momentum, i.e., ${\cal F}^i p_i =0$.
This four-vector can be modelled as a sum of two forces
\begin{equation}\label{force1}
{\cal F}^i = \frac{g}{mc} Q^{(a)} F_{(a)}^{ik} \ p_k + G^i_{({\rm
self})} \,.
\end{equation}
The first part is clearly the non-Abelian generalization of the
Lorentz force in $U(1)$ - electrodynamics, and $F^{ik}_{(a)}$ is
the strength of the Yang-Mills field. The second part includes the
self-interaction forces of different types. We will discuss their
structure below. Analogously, due to (\ref{distrib}) the following
relations hold:
\begin{equation}\label{CharEq2}
G_{(a)(b)}{\cal H}^{(a)} Q^{(b)}=0 \,, \quad d_{(a)(b)(c)} H^{(a)}
Q^{(b)} Q^{(c)} = 0 \,, ...
\end{equation}
giving the key for modeling of the functions $H^{(a)}$.

\subsection{Macroscopic moments and balance equations}

The standard procedure of subsequent integration of the kinetic
equation yields the balance equations. The first one is the
conservation law for the particle number:
\begin{equation}\label{N}
\nabla_i N^i(x)=0 \,, \quad N^i(x)\equiv\int{{dP} d {\cal Q} \ f \
p^i}\,, \quad dP \equiv \sqrt{-g} \  d^4 p \,.
\end{equation}
The second balance equation
\begin{equation}\label{ConsT}
\nabla_k T^{ik}(x)=mc\int{{dP}{d {\cal Q}} \  f \ {\cal F}^i}\,,
\quad T^{ik}(x)\equiv \int{{dP} d {\cal Q} \ f \ p^i \ p^k}
\end{equation}
relates to the evolution of the particle stress-energy tensor
under the influence of the force ${\cal F}^i$. The equations
\begin{equation}\label{curr}
\nabla_k {\cal I}^{k (a)}(x)= \int{{dP}{d {\cal Q}} \  f \ {\cal
H}^{(a)}}\,, \quad {\cal I}^{k (a)}(x)\equiv \frac{1}{mc}
\int{{dP}{d {\cal Q}} \ Q^{(a)} p^k \ f }
\end{equation}
describe the evolution of the color currents ${\cal I}^{k (a)}$:
they become conservation laws if the integral of ${\cal H}^{(a)}$
vanishes. Finally, the balance of entropy is regulated by the
equation
\begin{equation}\label{EntTrans}
\sigma(x)\equiv\nabla_i{\cal S}^i= k_B mc^{2}\int{{dP} \ f
    \left[
        \frac{\partial{{\cal F}^i}}{\partial{p^i}}
              +\frac{\partial{{\cal H}^{(a)}}}{\partial{Q^{(a)}}} \right]} \,, \label{28}
\end{equation}
\begin{equation}\label{EntFlux}
{\cal S}^i(x)\equiv-k_B c\int{{dP}{d {\cal Q}} \  f \
p^i\left[\ln{h^*}f - 1\right]}\,,
\end{equation}
where $\sigma(x)$ is the entropy production scalar.

\section{Non-minimal field equations }

\subsection{Gauge field}

The Yang-Mills field strength ${\bf F}_{mn}$ and the Yang-Mills
potential ${\bf A}_i$ are considered to take values in the Lie
algebra of the gauge group $SU(n)$ \cite{Rubakov}:
\begin{equation}
{\bf F}_{mn} = - i {\cal G} {\bf t}_{(a)} F^{(a)}_{mn} \,, \quad
{\bf A}_m = - i {\cal G} {\bf t}_{(a)} A^{(a)}_m \,,
\label{represent}
\end{equation}
where the real fields $F^{(a)}_{mn}$ and $A^{(a)}_m$ are connected
as follows
\begin{equation}
F^{(a)}_{mn} {=} \nabla_m A^{(a)}_n {-} \nabla_n A^{(a)}_m + {\cal
G} f^{(a)}_{\ (b)(c)} A^{(b)}_m A^{(c)}_n \,. \label{Fmn}
\end{equation}
The tensor $F^{ik}_{(a)}$ satisfies the relation
\begin{equation}
\hat{D}_k F^{*ik}_{(a)} = 0 \,, \quad F^{*ik}_{(a)} \equiv
\frac{1}{2}\epsilon^{ikls} F_{ls (a)} \,, \label{dual}
\end{equation}
where $\epsilon^{ikls} = \frac{1}{\sqrt{-g}} E^{ikls}$ is the
Levi-Civita tensor, $E^{ikls}$ is the completely antisymmetric
symbol with $E^{0123}{=} 1$. The gauge invariant derivative
$\hat{D}_k$ is defined as
\begin{eqnarray}
\hat{D}_m T^{(a) \cdot \cdot \cdot}_{\cdot \cdot \cdot (d)} \equiv
\nabla_m T^{(a) \cdot \cdot \cdot}_{\cdot \cdot \cdot (d)} + {\cal
G} f^{(a)}_{\cdot (b)(c)} A^{(b)}_m T^{(c) \cdot \cdot
\cdot}_{\cdot \cdot \cdot (d)} - {\cal G} f^{(c)}_{\cdot (b)(d)}
A^{(b)}_m T^{(a) \cdot \cdot \cdot}_{\cdot \cdot \cdot (c)} +...
\,,  \label{DQ2}
\end{eqnarray}
where $T^{(a) \cdot \cdot \cdot}_{\cdot \cdot \cdot (d)}$ is
arbitrary tensor in the group space. The Yang-Mills field strength
$F^{ik}_{(a)}$ is considered to be a solution of the non-minimally
extended master equations for the gauge field. In the
three-parameter non-minimal Einstein-Yang-Mills model
\cite{1BZ06,2BZ06,BSZ07} these equations read
\begin{equation}
\hat{D}_k \left[ F^{ik(a)} + {\cal R}^{ikmn}F^{(a)}_{mn} \right] =
{\cal I}^{i(a)} \,, \label{Heqs}
\end{equation}
where the color current ${\cal I}^{i(a)}$ is given by
(\ref{curr}), and the susceptibility tensor ${\cal R}^{ikmn}$ is
\begin{equation}
{\cal R}^{ikmn} \equiv
\frac{q_1}{2}R\,(g^{im}g^{kn}{-}g^{in}g^{km}) {+}
\frac{q_2}{2}(R^{im}g^{kn} {-} R^{in}g^{km} {+} R^{kn}g^{im}
{-}R^{km}g^{in}) {+} q_3 R^{ikmn}\,. \label{sus}
\end{equation}
Here $q_1$, $q_2$ and $q_3$ are the constants of non-minimal
coupling of the gauge and gravity fields, $R^{ikmn}$ is the
Riemann tensor, $R^{ik}$ is the Ricci tensor, $R$ is the Ricci
scalar.

\subsection{Gravitational field}

In the three-parameter non-minimal theory, linear in curvature,
the equations for the gravity field  take the standard form (see
\cite{BL05,1BZ06,2BZ06,BSZ07})
\begin{equation}
\left(R_{ik}-\frac{1}{2}Rg_{ik}\right)= \Lambda g_{ik} +
{\kappa}\left(  T^{(YM)}_{ik} + T_{ik}  + T^{(NM)}_{ik} \right)
\,. \label{Eeq}
\end{equation}
The stress-energy tensor of pure Yang-Mills field is given by the
term $T^{(YM)}_{ik}$:
\begin{equation}
T^{(YM)}_{ik} \equiv \frac{1}{4} g_{ik} F^{(a)}_{mn}F^{mn}_{(a)} -
F^{(a)}_{in}F_{k\,(a)}^{\ n} \,. \label{TYM}
\end{equation}
The term $T_{ik}$ denotes a stress-energy tensor of color
particles, given by (\ref{ConsT}). The non-minimal contribution
$T^{(NM)}_{ik}$ is presented by the decomposition
\begin{equation}
T^{(NM)}_{ik} = q_1 T^{(I)}_{ik} + q_2 T^{(II)}_{ik} + q_3
T^{(III)}_{ik} \,,  \label{Tdecomp}
\end{equation}
\begin{equation}%
T^{(I)}_{ik} = R\,T^{(YM)}_{ik} -  \frac{1}{2} R_{ik}
F^{(a)}_{mn}F^{mn}_{(a)} + \frac{1}{2} \left[ {\hat{D}}_{i}
{\hat{D}}_{k} - g_{ik} {\hat{D}}^l {\hat{D}}_l \right]
\left[F^{(a)}_{mn}F^{mn}_{(a)} \right] \,, \label{TI}
\end{equation}%
\[%
T^{(II)}_{ik} = -\frac{1}{2}g_{ik}\biggl[{\hat{D}}_{m}
{\hat{D}}_{l}\left(F^{mn(a)}F^{l}_{\ n(a)}\right)-R_{lm}F^{mn (a)}
F^{l}_{\ n(a)} \biggr] -{} \]%
\[{}- F^{ln(a)}
\left(R_{il}F_{kn(a)} + R_{kl}F_{in(a)}\right)-R^{mn}F^{(a)}_{im}
F_{kn(a)} - \frac{1}{2} {\hat{D}}^m{\hat{D}}_m \left(F^{(a)}_{in}
F_{k\,(a)}^{ \
n}\right)+ {}\]%
\begin{equation}%
\quad{}+\frac{1}{2}{\hat{D}}_l \left[ {\hat{D}}_i \left(
F^{(a)}_{kn}F^{ln}_{(a)} \right) + {\hat{D}}_k
\left(F^{(a)}_{in}F^{ln}_{(a)} \right) \right] \,, \label{TII}
\end{equation}%
\[
T^{(III)}_{ik} = \frac{1}{4}g_{ik} R^{mnls}F^{(a)}_{mn}F_{ls(a)}-
\frac{3}{4} F^{ls(a)} \left(F_{i\,(a)}^{\ n} R_{knls} +
F_{k\,(a)}^{\ n}R_{inls}\right) -\]%
\begin{equation}%
\quad {}-\frac{1}{2}{\hat{D}}_{m} {\hat{D}}_{n} \left[ F_{i}^{ \ n
(a)}F_{k\,(a)}^{ \ m} + F_{k}^{ \ n(a)} F_{i\,(a)}^{ \ m} \right]
\,. \label{TIII}
\end{equation}%
The conservation law is valid for the total stress-energy tensor:
\begin{equation} \nabla^k \left(T^{(YM)}_{ik} + T_{ik}  + T^{(NM)}_{ik}\right)
=0\,,
\label{Eeeq}
\end{equation}
this fact can be verified directly, using the balance equations
and Bianchi identities.

\section{Non-minimal color forces}

We obtained a self-consistent non-minimally extended evolutionary
model for the system of colored particles, which interact by gauge
and gravitational fields. This model is clearly a non-Abelian
non-minimal generalization of the Einstein-Maxwell-Vlasov model,
which is investigated in detail by the Kazan Gravitational Group
in 1975-2005 (see historical review in this issue). The system is
non-linear and self-consistent. For instance, in order to find the
Yang-Mills field strength $F^{ik}_{(a)}$ we should solve the
equations (\ref{Heqs}), in which the color currents ${\cal
I}^{k(a)}$ are macroscopic moments of the distribution function
$f$. The latter is a solution of kinetic equation (\ref{GRKE}),
which contains, in its turn, the tensor $F^{ik}_{(a)}$, which we
search for. Thus, the kinetic equation becomes {\it implicitly}
non-minimal due to the curvature coupling of gravitational and
gauge fields. Let us now discuss a new aspect in the non-minimal
extension of the theory of non-Abelian plasma, namely, the {\it
explicit} non-minimal generalization of the force-like structures
appeared in the kinetic equation.

\subsection{Non-minimal self-force $G^i_{({\rm self})}$}

The simplest reconstruction of this self-force, orthogonal to the
particle four - momentum, yields
\begin{equation}
G^i_{({\rm self})} = \left( \lambda_1 R \ U^k + \lambda_2 R^k_m \
U^m \right)\left( \delta^i_k \ p_m p^m - p^i  p_k \right) +
\lambda_3 R^{i}_{ \ kmn} p^k U^m p^n + ...  \label{gforce}
\end{equation}
by analogy with the decomposition of the susceptibility tensor
(\ref{sus}). Here $U^k$ is a four-vector of macroscopic velocity
of the multi-particle colored system. This time-like four-vector
can be obtained by the Eckart receipt as $U^k = N^k /
\sqrt{N^mN_m}$. It is important that the self-force contains not
only the microscopic particle momentum $p^k$, but the macroscopic
quantity $U^k(x)$, as well. This explain the word
"self-interaction" in the force indication. If we take into
account the particle spin, the modeling possibilities  of this
force may be extended significantly (see, e.g.,
\cite{KuBa1,KuBa2}).

\subsection{Non-minimal term $H^{(a)}$}

We suggest the following non-minimal phenomenological
generalization of the Wong equations \cite{Wong}
\begin{equation}
\frac{\hat{D}}{ds} Q^{(a)} \equiv \frac{d}{ds} Q^{(a)} +
\frac{g}{mc} f^{(a)}_{\ (b)(c)} p^k A_k^{(b)} Q^{(c)} =
H^{(a)}_{({\rm NM})} \,. \label{gwong}
\end{equation}
In the minimal theory the term $H^{(a)}_{({\rm NM})}$ vanishes,
and we cover the standard case (see \cite{Else,Litim,BaSu,BKZ}).
The non-minimal term $H^{(a)}_{({\rm NM})}$ can be modelled along
the line of the decomposition of the susceptibility tensor
(\ref{sus}) as
\begin{equation}
H^{(a)}_{({\rm NM})} {=} \left[\delta^{(a)}_{(b)} {-}
\frac{Q^{(a)}Q_{(b)}}{Q^{(c)}Q_{(c)}} \right] F^{(b)}_{mn}\left[
\omega_1 R \ p^m U^n {+} \omega_2 R^m_s \left(p^nU^s {+}p^sU^n
\right) {+} \omega_3 R^{mn}_{\ \ \ ls}p^lU^s {+} ...\right]
\label{GHwong}
\end{equation}
The equation (\ref{gwong}) for the evolution of a color charge is
gauge-covariant and contains the susceptibility-like tensorial
structures, which are adopted in the non-minimal Einstein-Maxwell
theory \cite{qqq1}-\cite{qqq5}.

\vspace{5mm} \noindent {\bf To conclude} it is worth stressing
that the presented model of evolution of the non-minimal
non-Abelian plasma is so far a general scheme to be supplemented
with exact solutions. We hope to discuss the corresponding
solutions of cosmological type, solutions with pp-wave and
spherical symmetries in future papers.

\section*{Acknowledgments}
This work was supported by the Deutsche Forschungsgemeinschaft
through project No. 436RUS113/487/0-5.


\begin{thebibliography}{99}

\bibitem{Else} H.T. Elze and  U. Heinz.  {\it Phys. Rept.} (1989) {\bf 183}, 81.

\bibitem{Litim} D.F. Litim and C. Manuel. {\it Phys. Rept.} (2002) {\bf
364}, 451.

\bibitem{Hwa} R.C. Hwa  (Editor). {\it Quark-gluon plasma} (World
Scientific, Singapore, 1990).

\bibitem{FaraR} V. Faraoni, E. Gunzig and P. Nardone. {\it Fundamentals of Cosmic
Physics} (1999) {\bf 20}, 121.

\bibitem{Hehl3} F.W. Hehl and Yu.N. Obukhov.  {\it Lecture Notes in
Physics} (Springer, Berlin, 2001) {\bf 562}, 479.

\bibitem{MH} F. M\"uller-Hoissen. {\it Class. Quantum Grav.} (1988) {\bf 5}, L35.

\bibitem{Drum} I.T. Drummond and S.J. Hathrell.  {\it Phys. Rev.} (1980) {\bf D 22}, 343.

\bibitem{BL05} A.B. Balakin and J.P.S. Lemos. {\it Class. Quantum Grav.} (2005) {\bf 22}, 1867.

\bibitem{1BZ06} A.B. Balakin and A.E. Zayats. {\it Gravit. and Cosmol.} (2006) {\bf
12}, 302.

\bibitem{2BZ06} A.B. Balakin and A.E. Zayats. {\it Phys. Lett. B} (2007) {\bf 644}, 294.

\bibitem{BSZ07} A.B. Balakin, S.V. Sushkov and A.E. Zayats. {\it Phys. Rev.
D} (2007) {\bf 75}, 084042.

\bibitem{Laptev} B.L. Laptev. In: {\it Geometry and Theory of
Relativity} (Kazan University Press, Kazan, 1958) 75.

\bibitem{Vlasov} A.A. Vlasov {\it Statistical Distribution
Functions} (Nauka, Moscow, 1966).

\bibitem{Isr78}  W. Israel. {\it Gen. Rel. Grav.} (1978) {\bf 9}, 451.

\bibitem{FT80} Y. Feldman and G.E. Tauber. {\it Gen. Rel. Grav.} (1980) {\bf 12}, 837.

\bibitem{Krasnoyarsk} A.B. Balakin and V.R. Kurbanova. {\it Bull. Krasnoyarsk State University}
(2005), N 7, 30.

\bibitem{Wong} S.K. Wong. {\it Nuovo Cimento} (1970) {\bf 65 A}, 689.

\bibitem{Rubakov} V. Rubakov {\it Classical Theory of Gauge
Fields} (Princeton University Press, Princeton and Oxford, 2002).

\bibitem{KuBa1} A. Balakin, V. Kurbanova and W. Zimdahl. {\it Gravit. and Cosmol.} (2002) {\bf 8},
Suppl.II, 6.

\bibitem{KuBa2} A. Balakin and V. Kurbanova. {\it Gravit. and
Cosmol.} (2004) {\bf 10}, 98.

\bibitem{BaSu} A.B. Balakin  and  F.G. Suslikov.  {\it  Compt. Rend. Acad.
Sci. Paris} (1997) {\bf 324} Serie II b, 619.

\bibitem{BKZ} A.B. Balakin, V.R. Kurbanova and W. Zimdahl. {\it J. Math.
Phys.} (2003) {\bf 44}, 5120.

\bibitem{qqq1} A.B. Balakin.  {\it Class. Quantum Grav.} (1997) {\bf 14}, 2881.

\bibitem{qqq2} A.B. Balakin and J.P.S. Lemos. {\it Class. Quantum
Grav.} (2001) {\bf 18}, 941.

\bibitem{qqq3} A.B. Balakin, R. Kerner and  J.P.S. Lemos. {\it Class. Quantum
Grav.} (2001) {\bf 18}, 2217.

\bibitem{qqq4} A.B. Balakin and J.P.S. Lemos. {\it Class. Quantum
Grav.} (2002) {\bf 19}, 4897.

\bibitem{qqq5} A.B. Balakin and W. Zimdahl. {\it Phys. Rev. D} (2005) {\bf 71},  124014.



\end{thebibliography}
\end{document}